\documentclass[linenumber,mathlines]{aastex631}
\usepackage{color,amsmath,amssymb,graphicx,latexsym,upgreek}
\usepackage{threeparttable,multirow,txfonts,subfigure,booktabs,appendix}
\usepackage{accents,slashed,ulem}

\date{\today}
\received{XXX}
\revised{YYY}
\accepted{ZZZ}
\shorttitle{Slow diffusion is necessary to explain the $\gamma$-ray pulsar halos}
\shortauthors{Bao et al.}
\graphicspath{{./}{figs/}}

\begin{document}

\title{Slow diffusion is necessary to explain the $\gamma$-ray pulsar halos}

\author{Li-Zhuo Bao}
\affiliation{Key Laboratory of Particle Astrophysics, Institute of High Energy Physics, Chinese Academy of Sciences, Beijing 100049, China}
\affiliation{University of Chinese Academy of Sciences, Beijing 100049, China}

\author{Kun Fang}
\affiliation{Key Laboratory of Particle Astrophysics, Institute of High Energy Physics, Chinese Academy of Sciences, Beijing 100049, China}
\affiliation{TIANFU Cosmic Ray Research Center, Chengdu, 610000 Sichuan, China}

\author{Xiao-Jun Bi}
\email{bixj@ihep.ac.cn}
\affiliation{Key Laboratory of Particle Astrophysics, Institute of High Energy Physics, Chinese Academy of Sciences, Beijing 100049, China}
\affiliation{University of Chinese Academy of Sciences, Beijing 100049, China}
\affiliation{TIANFU Cosmic Ray Research Center, Chengdu, 610000 Sichuan, China}

\author{Sheng-Hao Wang}
\affiliation{Key Laboratory of Particle Astrophysics, Institute of High Energy Physics, Chinese Academy of Sciences, Beijing 100049, China}
\affiliation{University of Chinese Academy of Sciences, Beijing 100049, China}

\begin{abstract}
It was suggested that the $\gamma$-ray halo around Geminga might not be
interpreted by slow-diffusion. If the ballistic regime of
electron/positron propagation is considered, the Geminga halo may be explained
even with a large diffusion coefficient. In this work, we examine this effect by
taking the generalized J\"uttner propagator as the approximate relativistic
Green's function for diffusion and find that the morphology of the Geminga halo
can be marginally fitted in the fast-diffusion scenario. However, the recently
discovered $\gamma$-ray halo around PSR J0622$+$3749 at LHAASO cannot be explained by the
same effect and slow diffusion is the only solution. Furthermore, both the two
pulsar halos require a conversion efficiency from the pulsar spin-down energy to
the high energy electrons/positrons much larger than 100\%, if they are interpreted
by this ballistic transport effect. Therefore, we conclude that slow diffusion is
necessary to account for the $\gamma$-ray halos around pulsars.
\end{abstract}

\section{Introduction}
\label{sec:intro}
It has been predicted that a middle-aged pulsar should be surrounded by a
$\gamma$-ray halo \citep{Linden:2017vvb}. The $\gamma$-ray halo is generated by
the high-energy electrons and positrons\footnote{For simplicity, we use
\textit{electrons} to denote both electrons and positrons hereafter.},
accelerated by the central pulsar wind nebula (PWN), scattering with the
background photons when they are injected and propagate in the interstellar
medium. The $\gamma$-ray halos around Geminga and Monogem were first detected by
the HAWC Collaboration \citep{Abeysekara:2017old}. The most intriguing feature of
the halos is that the diffusion coefficients indicated by the $\gamma$-ray
profiles are much smaller than the average value in the Galaxy derived by the B/C
data.

Recently, a paper by \citet{Recchia:2021kty} points out another plausible
scenario, in which the pulsar halo may be explained without a slow-diffusion
environment. It is well known that the non-relativistic diffusion equation has
the superluminal problem. For the relativistic particles injected within the time
$t_{\rm crit}\equiv3D/c^2$, the typical diffusion distance
$\lambda \sim \sqrt{Dt}$ is greater than $ct$, where $D$ is the diffusion
coefficient and $c$ is the light speed. For $t\ll t_{\rm crit}$, the particles
should propagate ballistically rather than diffusively. They found that the
ballistic propagation and the transition to the diffusion regime may explain the
Geminga halo even assuming a typical diffusion coefficient in the Galaxy.

The superluminal problem in non-relativistic diffusion equation 
has been studied by \citet{Aloisio:2008tx}. It is found
that the generalized J\"uttner propagator can be taken 
as the approximate relativistic Green's function for diffusion although a full 
relativistic diffusion equation has not been developed \citep{Aloisio:2008tx}.
The method used by \citet{Recchia:2021kty} can be considered as an
approximate form of the J\"uttner propagator to describe the ballistic and
quasi-ballistic propagation. 

In this work, we examine the effect of relativistic diffusion on the
interpretation of the $\gamma$-ray pulsar halos. We adopt the generalized
J\"uttner propagator as the relativistic Green's function for the diffusion
equation. Besides the Geminga halo that has been paid great attention, we also
test another newly discovered $\gamma$-ray halo around PSR J0622$+$3749 by
LHAASO-KM2A \citep{Aharonian2021extended}, which is very likely an analog of the Geminga halo. 

We find that if the effect of ballistic propagation is taken into account
the profile of the $\gamma$-ray halo around Geminga can be roughly fitted
and the result by \citet{Recchia:2021kty} is repeated.
However, the same effect 
cannot account for the $\gamma$-ray halo profile around PSR J0622$+$3749
and slow diffusion is still necessary in this case.
Furthermore, for the large diffusion coefficients the conversion
efficiencies from pulsar spin-down energy to the high energy electrons 
are much greater than 100\% for both the two halos. 

%the energy conversion efficiency, which is defined as the ratio between the total energy of high energy electrons and positrons to the pulsar spin down energy, increases. We found it is generally greater than 100\% for a large diffusion coefficient. Therefore, the scenario that the halo is due to the ballistic propagation in a large diffusion coefficient is strongly disfavored.

In the next section, we will present our calculation and the results.
Then we give the conclusion.

\section{Calculation and Results}
\label{sec:cal&res}
Pulsar halos are generated by electrons escaping from the PWNe and wandering in
the ISM. Thus, solving the electron propagation equation in the ISM is the core
of calculating the halo morphology. The electron propagation can be expressed by
the diffusion-cooling equation:
\begin{equation}
\frac{\partial N(E, \boldsymbol{r}, t)}{\partial t} = D(E)\Delta N(E, \boldsymbol{r}, t)+\frac{\partial[b(E)N(E, \boldsymbol{r}, t)]}{\partial E} + Q(E, \boldsymbol{r}, t)\ ,
\label{eq:prop}
\end{equation}
where $E$ is the electron energy, and $N$ is the electron number density. The
diffusion coefficient is taken as $D(E)=D_0(E/1~{\rm GeV})^\delta$, where we
assume $\delta=1/3$ following the Komolgorov's theory.

The energy-loss rate is defined by $b(E) = -{\rm d}E/{\rm d}t$ and can be
expressed by $b(E) = b_0E^2$. ICS and synchrotron radiation dominate the energy
losses of high-energy electrons. Considering the Klein-Nishina effect in ICS,
$b_0$ should be energy dependent. We adopt the parameterization given by
\citet{Fang:2020dmi} to precisely calculate the ICS component. The seed
photon fields for ICS are taken from \citet{Abeysekara:2017old} and
\citet{Fang:2021qon} for Geminga and LHAASO J0621$+$3755, respectively. We take
the magnetic field to be 3~${\rm \mu}$G for both the cases to get the
synchrotron component. 

The source function is taken as
\begin{equation}
Q(E,\boldsymbol{r},t)=\left\{
\begin{aligned}
& q(E)\ \delta(\boldsymbol{r}-\boldsymbol{r}_s)\ (1+t/t_{\rm sd})^{-2}/(1+t_s/t_{\rm sd})^{-2}\ , & t\geq0 \\
& 0\ , & t<0
\end{aligned}
\right.\ ,
\label{eq:src}
\end{equation}
where $q(E)$ is the electron injection spectrum at the current time,
$\boldsymbol{r}_s$ is the position of the pulsar, $t_s$ is the pulsar age, and
$t_{\rm sd}$ is the typical spin-down time scale of pulsar,
which is set to be 10 kyr.

We assume the injection spectrum to be a power-law form with an
exponential cutoff:
\begin{equation}
q(E)=q_0(E/{\rm 1~GeV})^{-p}\ {\rm exp}[-(E/E_c)^2]\ ,
\label{eq:inj}
\end{equation}
where the values of $p$ and $E_c$ are listed in Table.~\ref{table:parameters}, and
the normalization $q_0$ can be determined by the relation
\begin{equation}
\int q(E)E{\rm d}E=\eta L\ ,
\label{eq:eta}
\end{equation}
where $L$ is the current spin-down luminosity of the pulsar and $\eta$ is the
conversion efficiency from the spin-down energy to the electron energy. For the
Geminga halo, the spectral parameters are determined by a rough fit to a
preliminary $\gamma$-ray spectrum of HAWC \citep{2019ICRC...36..832Z}. For LHAASO
J0621$+$3755, we adopt the best-fit spectral parameters to the LHAASO-KM2A
spectrum \citep{Aharonian2021extended,Fang:2021qon}.

We can obtain the solution of Eq.~(\ref{eq:prop}) with the Green's function
method:
\begin{equation}
N(E, \boldsymbol{r}, t) = \int_{\mathbb{R}^3}{\rm d}^3\boldsymbol{r}_0\int_{-\infty}^{t}{\rm d}t_0\int_{-\infty}^{+\infty}
{\rm d}E_0\,G(E, \boldsymbol{r}, t; E_0, \boldsymbol{r}_0, t_0)\,Q(E_0, \boldsymbol{r}_0, t_0)\ .
\label{eq:solution}
\end{equation}
In the normal diffusion model, the Green's function is expressed as
\begin{equation}
G(E, \boldsymbol{r}, t; E_0, \boldsymbol{r}_0,
t_0) = \frac{1}{b(E)(\uppi\lambda^2)^{3/2}}\ {\rm exp}\left[-\frac{(\boldsymbol{r}-\boldsymbol{r}_0)^2}{\lambda^2}\right]\ \delta(t-t_0-\tau)\ H(\tau)\ ,
\label{eq:green}
\end{equation}
where
\begin{equation}
\lambda = 2\sqrt{\int_{E}^{E_0}\ \frac{D(E^{\prime})}{b(E^{\prime})}\ {\rm d}E^{\prime}}\ ,\quad\tau=\int_E^{E_0}\frac{{\rm d}E'}{b(E')}\ ,
\end{equation}
and $H$ is the Heaviside step function.

In order to eliminate superluminal motion, we need to find the relativistic
correction of Eq.~(\ref{eq:prop}). Unfortunately, many efforts over decades have
not been successful so far. To evade this difficulty,
\citet{Dunkel:2007relativistic} noticed that the propagator
\begin{equation}
P(E, r, t) = \frac{1}{(\uppi\lambda^2)^{3/2}}{\rm exp}\left(-\frac{r^2}{\lambda^2}\right)
\label{eq:P_diff}
\end{equation}
has the same form with the Maxwell-Boltzmann speed
distribution of particles with mass $m$ in thermal equilibrium gas
with temperature $T$:
\begin{equation}
f_{\rm{M-B}}(\boldsymbol{v}) = \left(\frac{m}{2\uppi kT}\right)^{3/2}\ {\rm exp}\left(-\frac{mv^2}{2kT}\right)\ ,
\label{eq:distMB}
\end{equation}
if we make the replacements $v \rightarrow x$ and $2kT/m \rightarrow \lambda^2$.
\citet{Juttner1911maxwellsche} derived the Maxwell-J\"uttner distribution which
describes the speeds of relativistic particles in thermal equilibrium gas:
\begin{equation}
f_{\rm{M-J}}(\boldsymbol{p}) = \frac{1}{4\uppi (mc)^3}\ \frac{\mu}{K_2(\mu)}\ {\rm exp}(-\gamma \mu)\ ,
\end{equation}
where $\boldsymbol{p} = \gamma m \boldsymbol{v}$ represents the momentum,
$\gamma = 1/\sqrt{1-v^2/c^2}$ represents the Lorentz factor,
$\mu = mc^2/kT$, and $K_{\nu}$ is the $\nu$-order modified Bessel function of the
second kind. From the relationship
$4\uppi v^2f_v(\boldsymbol{v}){\rm d}v = 4\uppi p^2f_p(\boldsymbol{p}){\rm d}p$,
we derive
\begin{equation}
f_{\rm{M-J}}(\boldsymbol{v}) = \frac{\gamma^5}{4\uppi c^3}\ \frac{\mu}{K_2(\mu)}\ {\rm exp}(-\gamma \mu)\ .
\label{eq:distSMJ}
\end{equation}
However, the standard form Eq.~(\ref{eq:distSMJ}) might not represent the correct
relativistic equilibrium distribution \citep{dunkel2007relative}. A modification
\begin{equation}
f_{\rm{M-J}}(\boldsymbol{v}) = \frac{\gamma^4}{4\uppi c^3}\ \frac{\mu}{K_1(\mu)}\ {\rm exp}(-\gamma \mu)\ 
\label{eq:distMJ}
\end{equation}
is most frequently suggested.

It reminds us to take Eq.~(\ref{eq:distMJ}) with the same replacements
$v \rightarrow r$, $2kT/m \rightarrow \lambda^2$ and $c \rightarrow ct$
as the propagator in the Green's function, called the generalized J\"uttner
propagator
\citep{Aloisio:2008tx}:
\begin{equation}
P_{\rm rela}(E,r,t) = \frac{1}{4\uppi(ct)^3}\ \frac{H[1-\xi(r,t)]}{[1-\xi^2(r,t)]^2}\ \frac{\kappa(E,t)}{K_1[\kappa(E,t)]}\ {\rm exp}\left[-\frac{\kappa(E,t)}{\sqrt{1-\xi^2(r,t)}}\right]\ ,
\label{eq:P_rela}
\end{equation}
where $\xi(r,t) = r/ct$ and $\kappa(E,t) = 2(ct/\lambda)^2$. The
Heaviside step function $H$ is used to ensure that $P_{\rm rela}$ vanishes
when $r>ct$. In the extreme case of $\kappa \ll 1$ or $\kappa \gg 1$, the
propagator will transit to ballistic or diffusive propagator separately
(see Appendix \ref{sec:discuss}). In this way we derive the Green's function with
relativistic correction:
\begin{equation}
G_{\rm rela}(E, \boldsymbol{r}, t; E_0, \boldsymbol{r}_0,
t_0) = \frac{1}{4\uppi [c(t-t_0)]^3b(E)}\ \frac{H(1-\xi)}{(1-\xi^2)^2}\ \frac{\kappa}{K_1(\kappa)}\ {\rm exp}\left(-\frac{\kappa}{\sqrt{1-\xi^2}}\right)\ \delta(t-t_0-\tau)\ H(\tau)\ ,
\end{equation}
where $\xi = (\boldsymbol{r}-\boldsymbol{r}_0)/c(t-t_0)$
and $\kappa = 2[c(t-t_0)/\lambda]^2$.

\begin{figure}
	\centering
	\includegraphics[width=10cm]{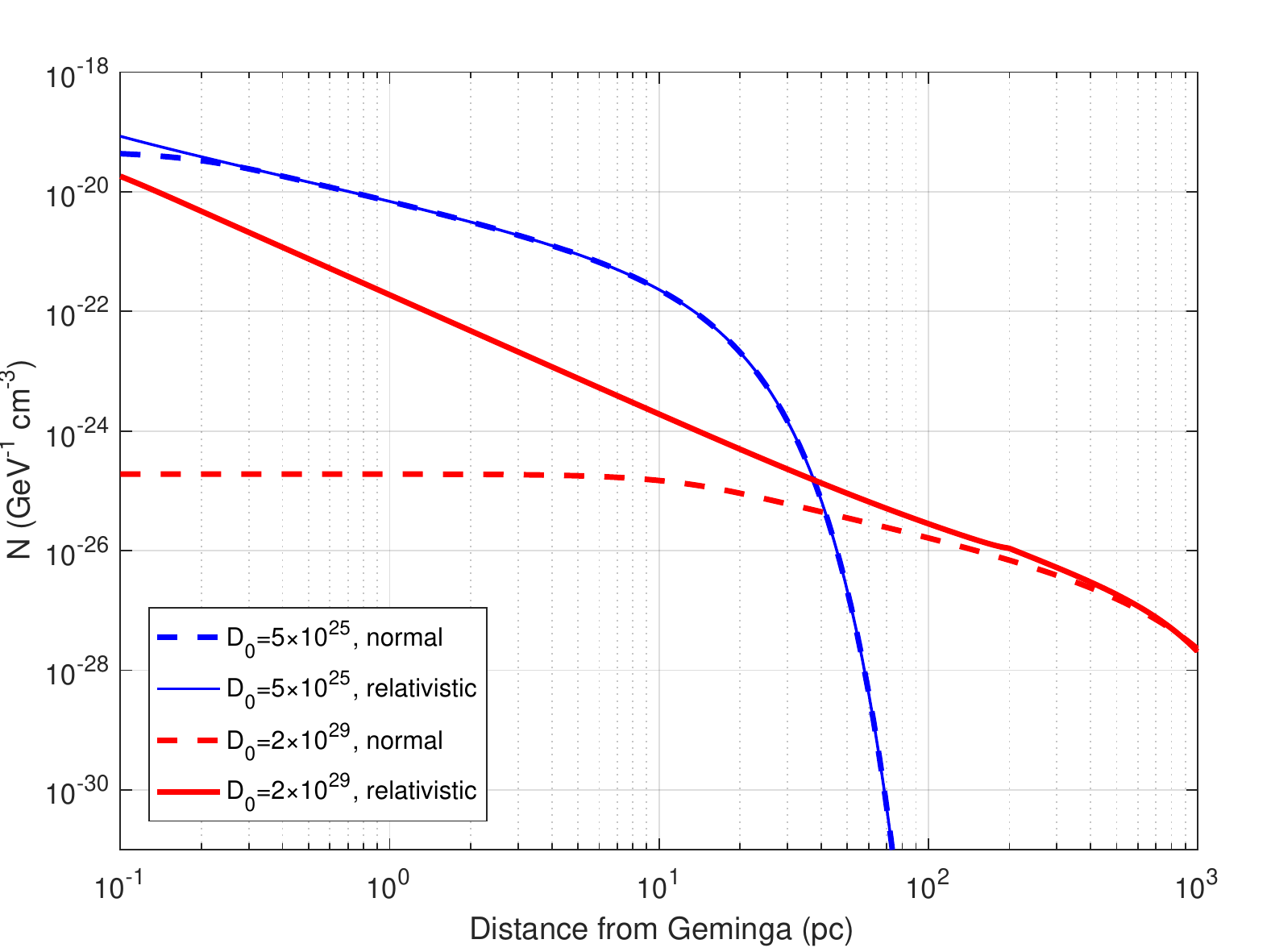}
	\caption{Electron number densities as functions of the distance from Geminga for the slow-diffusion ($D_0 = 5\times10^{25}$ cm$^2$ s$^{-1}$) and fast-diffusion ($D_0 = 2\times10^{29}$ cm$^2$ s$^{-1}$) scenarios. Both the relativistic and non-relativistic diffusion models are presented. The conversion efficiency is set as 100\% for all the cases in this figure.}
	\label{fig:density}
\end{figure}

In Fig.~\ref{fig:density}, we draw $N$ as functions of the distance from Geminga
to show the effect of relativistic correction. The electron energy is 100~TeV,
which is roughly the parent electron energy of the TeV pulsar halos. The
conversion efficiency is set as 100\% for all the cases. As expected, there is
little difference between the standard and relativistic diffusion models in the
slow-diffusion scenario. When the diffusion coefficient is significantly larger,
the relativistic correction is crucial. The newly injected electrons are still in
the ballistic regime and cannot flee far away from the source with superluminal
velocities, leading to a more constrictive distribution than that of the
non-relativistic case. However, although the fast-diffusion scenario can also
predict a steep electron distribution near the source after the relativistic
correction, the absolute $N$ is about an order of magnitude smaller than that of
the slow-diffusion scenario. It means that a significantly larger conversion
efficiency is required for the former to explain a same observation. 

Due to the anisotropy of inverse Compton scattering (ICS) processes of the
quasi-ballistic electrons within the vicinity of source, it is necessary to consider the
velocity angular distribution of the electrons in the small-angle region
with the following form \citep{Prosekin:2015}, the details of which are discussed in Appendix \ref{sec:M_alpha}:
\begin{equation}
M(x,\alpha) = \frac{1}{Z(x)}\ \mathrm{exp}[-\frac{3(1-\alpha)}{x}]\ ,
\label{eq:M_alpha}
\end{equation}
where $x(r,E)=rc/D(E)$, $Z(x)=\frac{x}{3}[1-\mathrm{exp}(-\frac{6}{x})]$ is the normalization
coefficient, and $\alpha$ is the cosine of the angle $\phi$ between the radial direction
and the line of sight.

We integrate $2NM$ over the line of sight from the Earth to an angle of $\theta$
observed away from the pulsar to get the apparent electron surface density and then obtain
the $\gamma$-ray surface brightness profile (SBP) with the standard calculation
of ICS \citep{Blumenthal:1970gc}. For the case of LHAASO J0621$+$3755, the
point-spread function (PSF) must be convoluted with our calculation results:
\begin{equation}
\frac{{\rm d}\Phi}{{\rm d}\Omega}^{\ast}(\theta_x, \theta_y) = (\frac{{\rm d}\Phi}{{\rm d}\Omega} \ast {\rm PSF})\ (\theta_x, \theta_y) = \iint_{\mathbb{R}^2} \frac{{\rm d}\Phi}{{\rm d}\Omega}(\delta_x, \delta_y)\ {\rm PSF}(\theta_x-\delta_x, \theta_y-\delta_y)\ {\rm d}\delta_x\ {\rm d}\delta_y\ ,
\label{eq:conv}
\end{equation}
where ${\rm d}\Phi/{\rm d}\Omega$ is the flux per unit solid angle at the point
with coordinates $(\theta_x, \theta_y)$ in the image plane predicted by our model,
the star mark ${\ast}$ represents the convolution results. We emphasize that the PSF
is a 2-dimentional (2D) Gaussian function as
\begin{equation}
{\rm PSF}(\theta_x, \theta_y) = \frac{1}{2\uppi \sigma^2}\ {\rm exp}\left(-\frac{\theta_x^2+\theta_y^2}{2\sigma^2}\right)\ ,
\label{eq:PSF}
\end{equation}
where $\sigma = 0.45^\circ$ \citep{Aharonian2021extended}.
For each angular distance $\theta$, we take the average value of
$({\rm d}\Phi/{\rm d}\Omega)^{\ast}$ to get the SBP:
\begin{equation}
{\rm SBP}(\theta) = \frac{1}{2\uppi}\ \int_{0}^{2\uppi}\frac{{\rm d}\Phi}{{\rm d}\Omega}^{\ast}(\theta{\rm cos}\varphi, \theta{\rm sin}\varphi)\ {\rm d}\varphi\ .
\end{equation}

We vary the diffusion coefficient $D_0$ from $10^{25}$ to $10^{30}$ cm$^2$
s$^{-1}$ and take the energy conversion efficiency $\eta$ as the free parameter
to fit the HAWC data for Geminga and the LHASSO-KM2A data for LHAASO
J0621$+$3755, using the least-$\chi^2$ method. Other determined parameters are
summarized in Table.~\ref{table:parameters}. The details about the parameters 
$E_c$ and $p$ of the injection spectrum refer to Appendix \ref{sec:p_eta}.

\begin{table}[htbp]
\caption{Model parameters for Geminga and LHAASO J0621$+$3755.}
\centering
\begin{tabular}{c c c}
\toprule
   & Geminga & J0621\\
\midrule
  $t_s~({\rm kyr})$ & $342$ & $208$\\
  $r_s~({\rm pc})$ & $250$ & $1600$\\
  $L~(\rm{erg~s^{-1}})~$ & $~3.2\times10^{34}~$ & $~2.7\times10^{34}$\\
  $p$ & $1.0$ & $1.5$\\
  $E_c~({\rm TeV})$ & $133$ & $265$\\
\bottomrule
\end{tabular}
\label{table:parameters}
\end{table}

\begin{figure}
	\centering
	\includegraphics[width=8cm]{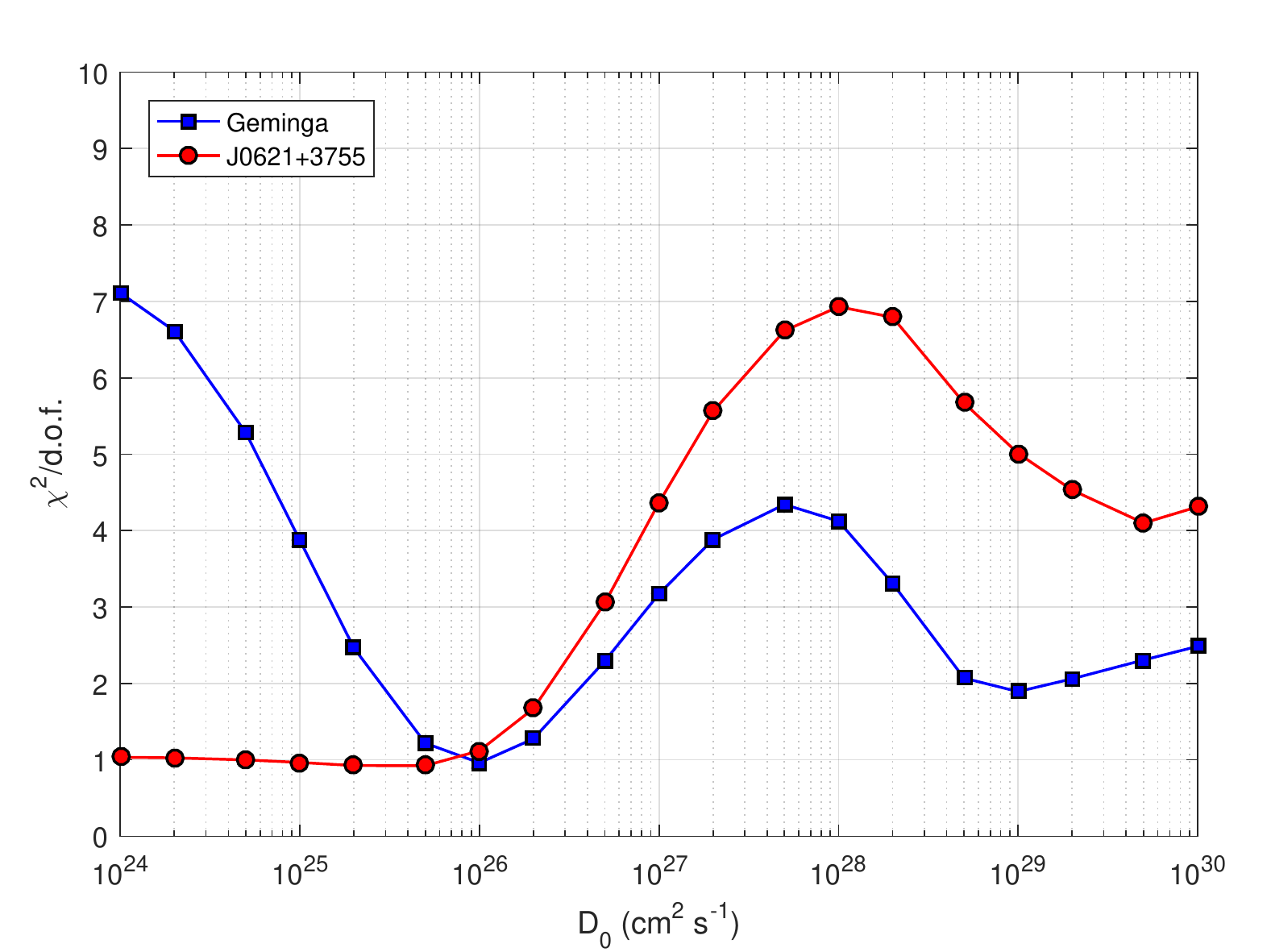}
	\includegraphics[width=8cm]{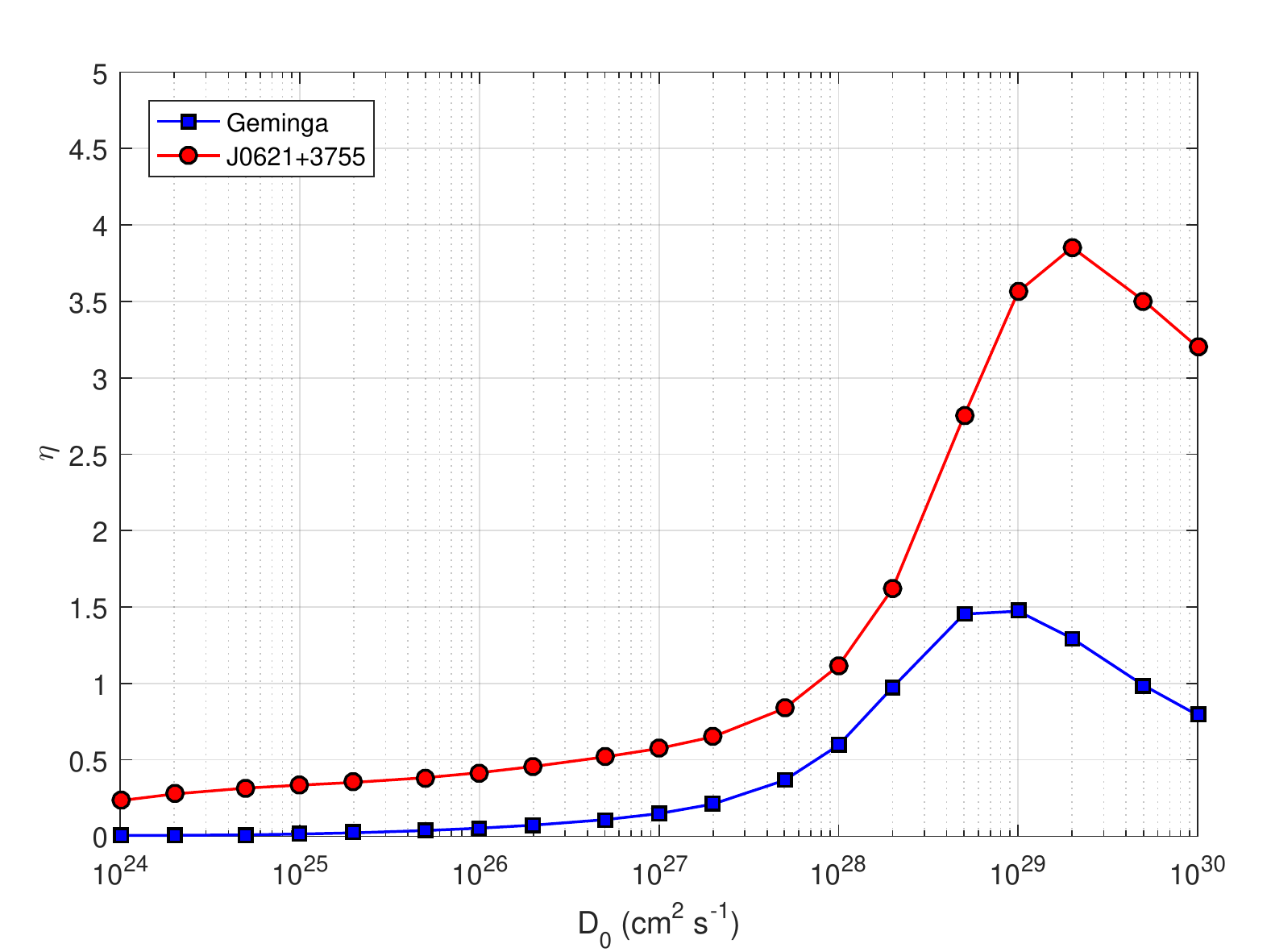}
	\caption{Left: The minimal $\chi^2$ of best fitting to the $\gamma$-ray halo profiles as function of diffusion coefficient at 1 GeV. Right: The efficiency of energy transfer from the pulsar spin-down energy to electron/positron pairs for different values of $D_0$.}
	\label{fig:chi2}
\end{figure}

We illustrate the fitting results in Fig.~\ref{fig:chi2}. It is clear that there
are two minimal values of $\chi^2$ for the Geminga case. The solution of the slow-diffusion scenario with
$D_0=1\times10^{26}~\mathrm{cm^2~s^{-1}}~[D(100~\mathrm{TeV})=5\times10^{27}~\mathrm{cm^2~s^{-1}}]$
gives very good fit to the data with $\chi^2/\rm{d.o.f.} \sim 1$. The fast-diffusion scenario with
$D_0=1\times10^{29}~\mathrm{cm^2~s^{-1}}~[D(100~\mathrm{TeV})=5\times10^{30}~\mathrm{cm^2~s^{-1}}]$
gives a poorer fit with $\chi^2/\rm{d.o.f.} \sim 2$, which is in agreement with the result
by \citet{Recchia:2021kty}. The best fitted fast diffusion coefficient is slightly larger than the Galactic typical
value derived by the B/C data \citep{Yuan:2017ozr}. The required $\eta$ for the slow-diffusion scenario is
$\sim 5\%$, while the fast-diffusion scenario needs a conversion efficiency of
$\sim150\%$, exceeding 100\%.

There are also two local minimal $\chi^2$ for the case of LHAASO J0621$+$3755, corresponding to
$D_0=5\times10^{25}~\mathrm{cm^2~s^{-1}}~[D(100~\mathrm{TeV})=2\times10^{27}~\mathrm{cm^2~s^{-1}}]$ and
$D_0=5\times10^{29}~\mathrm{cm^2~s^{-1}}~[D(100~\mathrm{TeV})=2\times10^{31}~\mathrm{cm^2~s^{-1}}]$, respectively.
However, only the slow-diffusion solution can fit the LHAASO-KM2A data. The
minimal $\chi^2/\rm{d.o.f.}$ in the fast-diffusion regime is about 4,
corresponding an exclusion with a confidence level of 99.996\%. The problem of
conversion efficiency is also serious for the fast-diffusion solution,
as $\eta \sim350\%$. This value may not be reasonable even considering the uncertainties
of the pulsar spin-down luminosity and the required total electron energy.

\begin{figure}
	\centering
	\includegraphics[width=8cm]{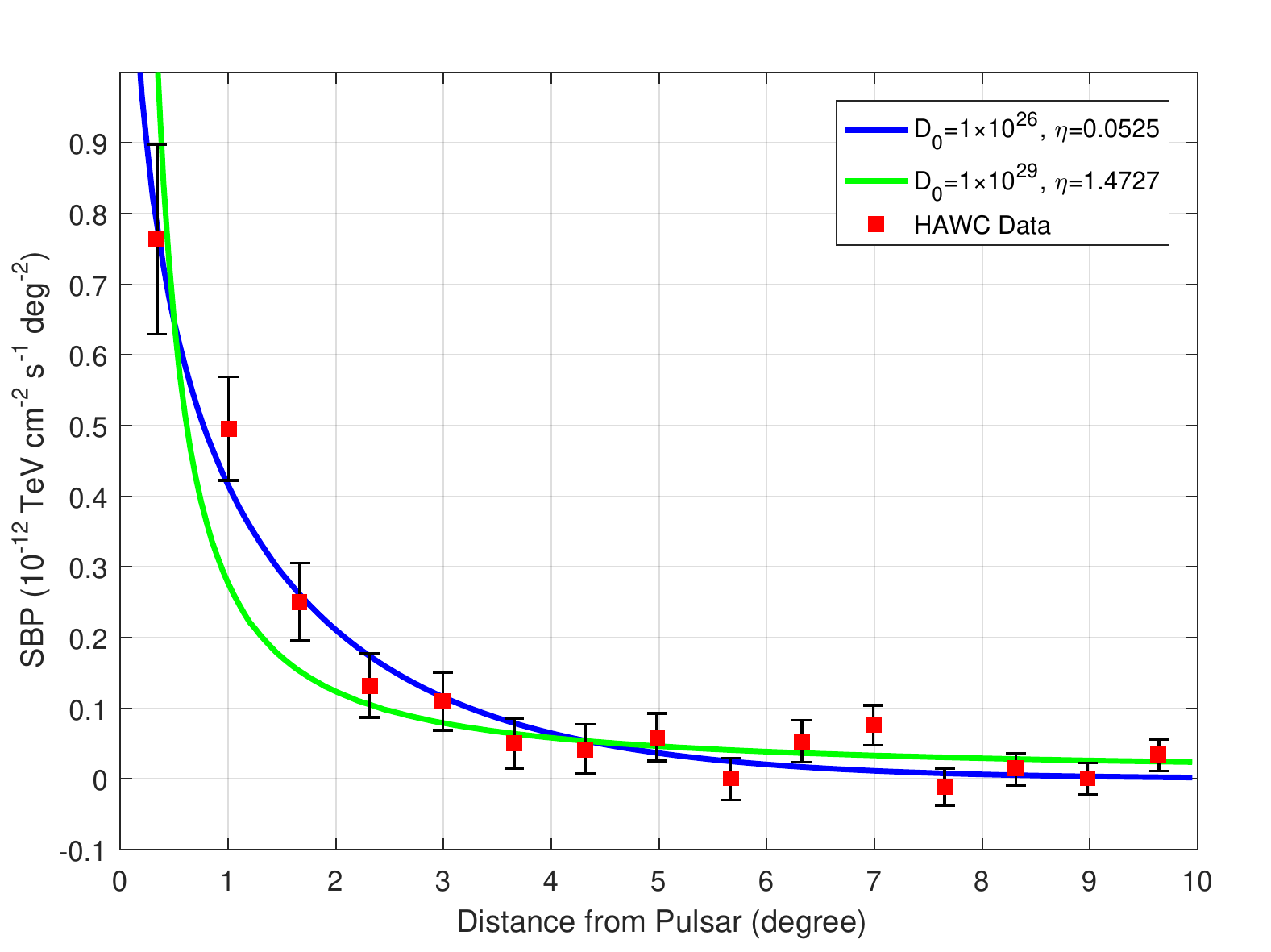}
    \includegraphics[width=8cm]{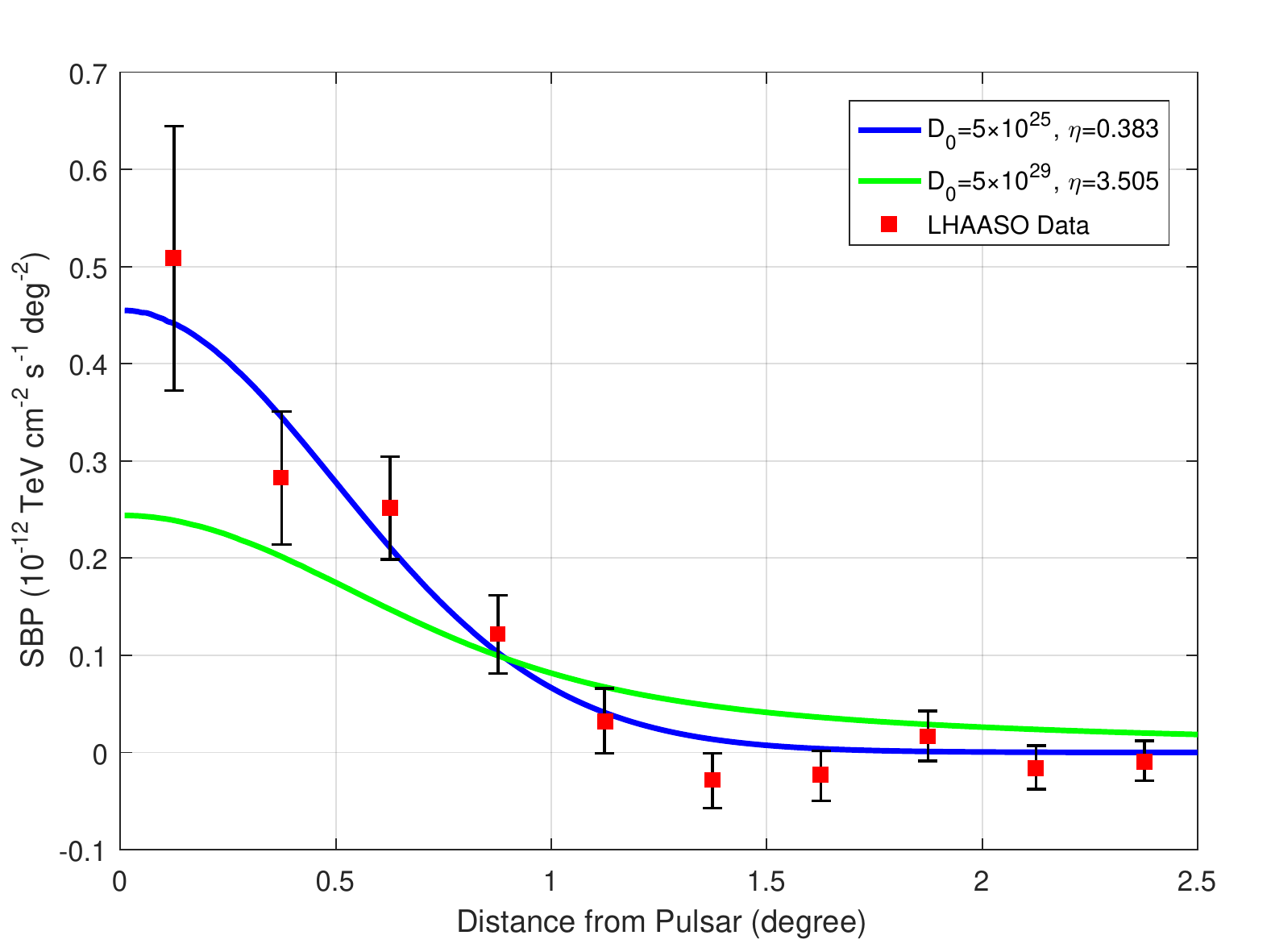}
	\caption{ The $\gamma$-ray profiles of the Geminga halo (left) and LHAASO J0621$+$3755 (right) taking $D_0$ corresponding to the two local minimal $\chi^2$.}
	\label{fig:profile}
\end{figure}

In Fig.~\ref{fig:profile}, we show the theoretical SBPs corresponding to the two
local minimal $\chi^2$ for both the Geminga halo and LHAASO J0621$+$3755. As
already indicated by the $\chi^2$ test, the fast-diffusion scenario can roughly
fit the measured SBP of the Geminga halo, while gives a very poor fit to that of
LHAASO J0621$+$3755. The predicted SBP is systematically lower than the data
within $\sim1^\circ$ from the pulsar and higher outside.

\begin{figure}
	\centering
	\includegraphics[width=8.6cm]{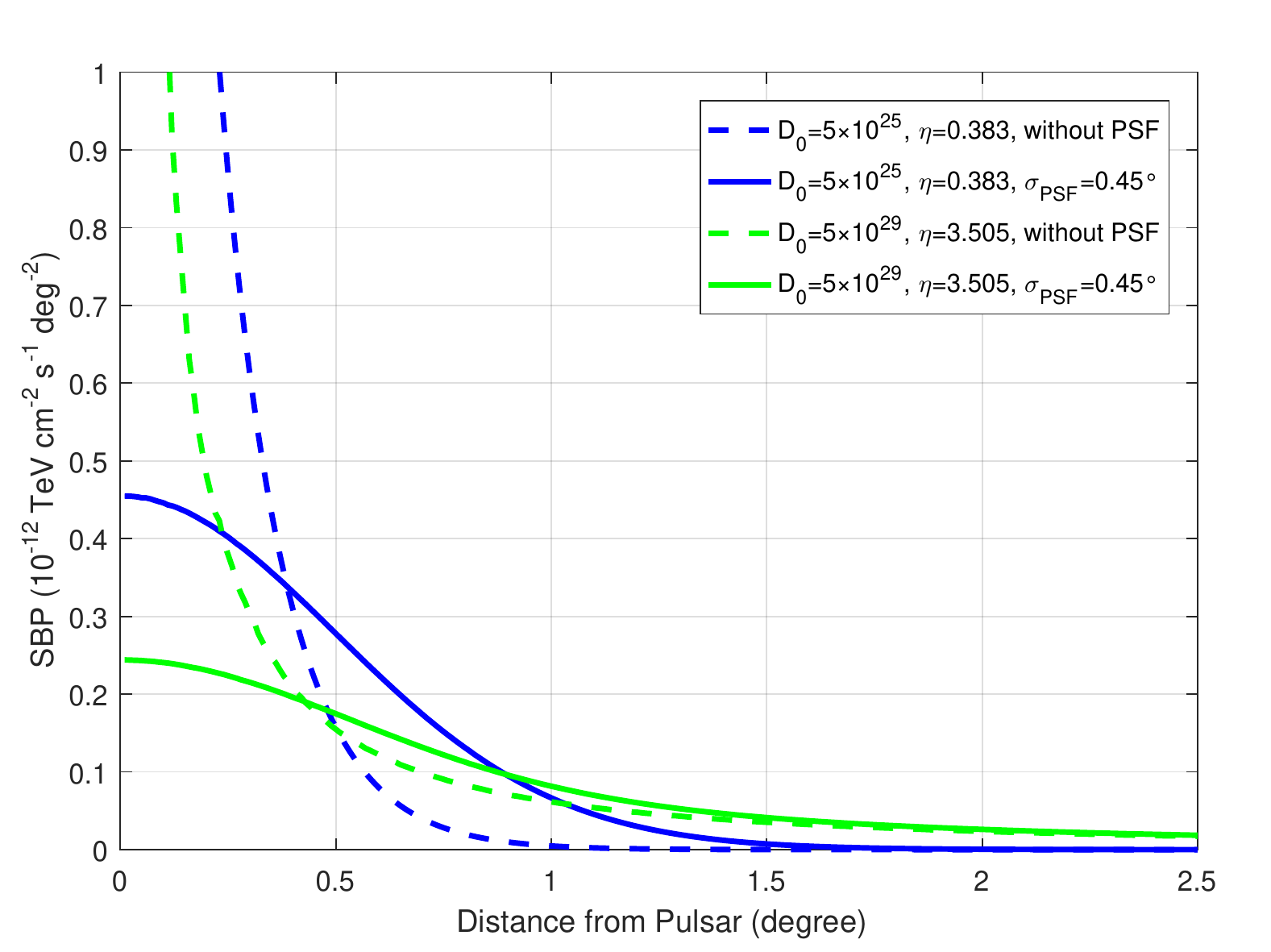}
	\caption{ Comparison of the $\gamma$-ray profiles for LHAASO J0621$+$3755 before and after PSF convolution. }
	\label{fig:psf}
\end{figure}

To understand the reason that the $\gamma$-ray profile of LHAASO J0621$+$3755
cannot be well fitted, we show the $\gamma$-ray profiles for LHAASO J0621$+$3755
with/without PSF convolution in solid/dashed lines in Fig.~\ref{fig:psf}.
For both slow-diffusion and fast-diffusion scenarios, $\eta$ is set to be the same value
of SBP blurred by a 2D Gaussian function with a size of 0.45$^\circ$ (in solid lines).
The trends of the two dashed lines are similar to those of Geminga.
In the case of fast diffusion, the green dashed line decreases more steeply in a small
region around the pulsar than the blue dashed line.
However, when the $\gamma$-ray profile is convolved with a large PSF,
each flux point is affected by a broad range of the original profile.
For example, the flux at $\theta=\sigma_{\rm PSF}=0.45^\circ$ of the original
profile contributes to the flux at $\theta=0^\circ$ of the convolved profile.
The original profile of the fast-diffusion case is significantly lower and flatter
than the slow-diffusion case when $\theta>0.1^\circ$. Thus, even though the original
profile of the fast-diffusion case is steeper in the most inner region, the convolved
profile is lower and flatter as affected by the flux of large angular distances.
Therefore, the green solid line becomes flatter after the PSF convolution and cannot fit
the data.

Moreover, the fast-diffusion scenario is \textit{not} equal to the pure
ballistic propagation, but should be seen as a combination of the ballistic
propagation for the latest injected electrons and the standard diffusion for early
injected electrons. The ballistic component originates from the freshly-injected
electrons with $t-t_0 \ll t_{\rm crit}$, and the diffusive component originates from
the electrons injected earlier with $t-t_0 \gg t_{\rm crit}$. The pure ballistic
component has a steep profile near the source and is a point-like source. The PSF
convolved profile of this component should be similar to the PSF. However, the
diffusive component is very extended in the fast diffusion scenario, and the PSF
convolved profile is significantly broader than the PSF. Thus, the superposition of
these two components don't follow the PSF shape.

\begin{figure}
	\centering
	\includegraphics[width=8cm]{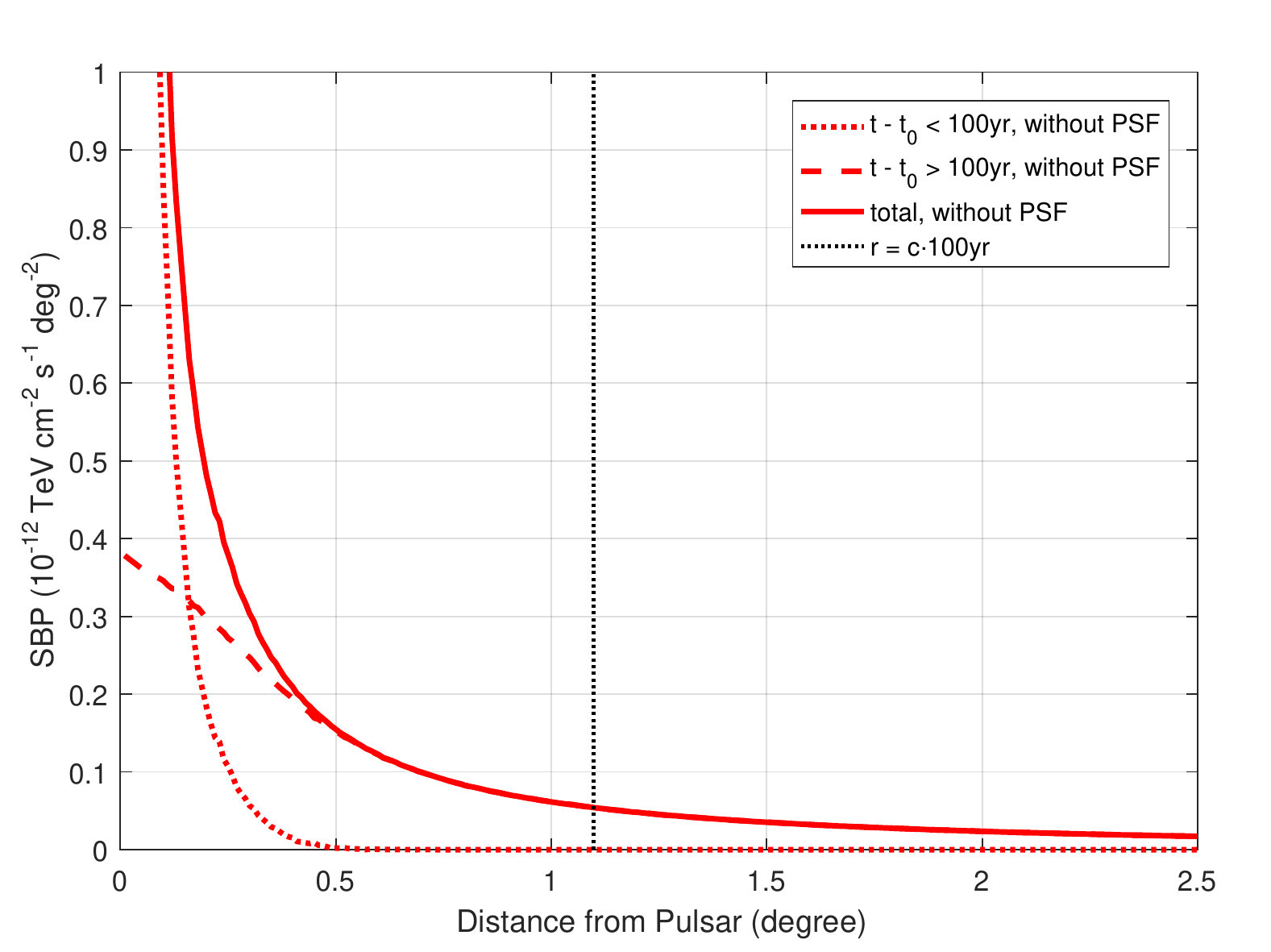}
    \includegraphics[width=8cm]{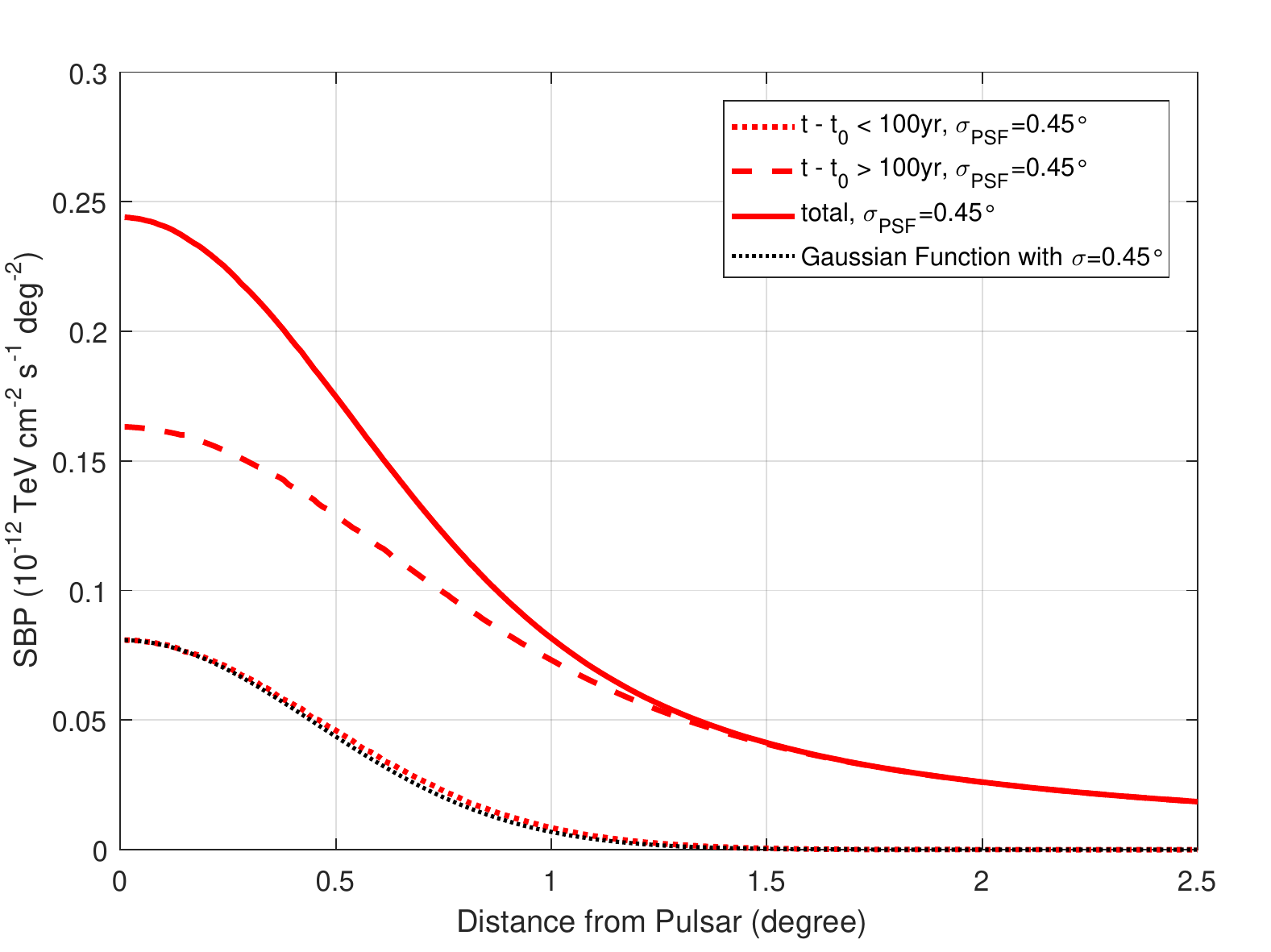}
	\caption{ The two parts of the $\gamma$-ray profile of LHAASO J0621$+$3755 under the fast-diffusion scenario without (left) and with (right) the PSF convolution.}
	\label{fig:recent}
\end{figure}

We illustrate this point in Fig.~\ref{fig:recent}. We divide the profile of
LHAASO J0621$+$3755 into two parts. One is generated by electrons injected in the
most recent 100~years, which is almost a pure ballistic regime. The other is
generated by electrons injected in the transitional (from pure ballistic to pure
diffusion) and diffusive regimes. The left panel of Fig.~\ref{fig:recent} shows the
profiles without the PSF convolution. The ballistic component shown in red solid line is
very sharp and point-like, restricted to a small range around the source by the vertical black
line, which represents the speed of light. The red dashed line dominates the flux at the
outer region, consisting of a flat background contributed by the transitional and diffusive component.
%and a blunt peak at the intersection point with the vertical black line, contributed by the transition component.
Therefore, the total profile can be seen as a superposition of a point-like source and a disk-like source.
After the convolution with the PSF, the point-like component is significantly flattened and
has a similar profile with the PSF, as shown in the right panel of Fig.~\ref{fig:recent}.
Meanwhile, the disk-like component is less affected by the convolution as it is much broader
than the PSF. The total profile is the superposition of a PSF-like component and a dominant
component with a much larger extension, which is not a PSF shape.

It can also be seen in Fig.~\ref{fig:psf} that the profile larger than $0.5^\circ$
is affected little by the convolution for the fast-diffusion case. If we force
the inner fluxes to fit the data, the outer fluxes will be significantly higher
than the data. Thus, the fast-diffusion model cannot fit the observation for
the case of LHAASO J0621$+$3755.

\iffalse
\textbf{There is a severe conflict on the fitting of LHAASO J0621$+$3755 between
\citet{Recchia:2021kty} and our results. They find a best-fit value at $D(100~\rm{TeV})
\sim 0.6 \mbox{--} 6.4 \times 10^{30}$ cm$^2$ s$^{-1}$ for the fast-diffusion scenario. The goodness of fit is almost
equal to the slow-diffusion scenario, and the efficiency of energy conversion is between
40\% and 100\%. We completely follow their calculation steps, and realize that they apply
a wrong convolution method. They regard the PSF simply as an 1D Gaussian function,
and directly convolute it with their calculation results. But actually, the PSF is defined
in the 2D image plane. Once the convolution method and PSF form are replaced with
the correct one, as shown in Eq.~(\ref{eq:conv}) and Eq.~(\ref{eq:PSF}), the results for the
fast-diffusion scenario become similar to our conclusion.}
\fi

Our fitting result of LHAASO J0621$+$3755 is in conflict with that of \citet{Recchia:2021kty}, which finds that
the goodness of fit  for the fast-diffusion scenario is almost equal to the slow-diffusion scenario, and the efficiency of energy conversion is between
40\% and 100\%. 
The results are due to a wrong PSF convolution method that taking the PSF as an 1D Gaussian function in \citet{Recchia:2021kty}.
Actually, the PSF is defined
in the 2D image plane. Once the convolution method and the PSF form are replaced with
the correct one, as shown in Eq.~(\ref{eq:conv}) and Eq.~(\ref{eq:PSF}), the results for the
fast-diffusion scenario become nearly the same as ours.

\section{Conclusion}
\label{sec:conclude}
With a relativistic correction to the electron propagation equation, we examine
if the morphologies of pulsar halos can be explained under the fast diffusion
scenario, i.e., the typical diffusion coefficient in the Galaxy. 
%Our criteria are
%whether the measured $\gamma$-ray profiles can be well reproduced and the
%required conversion efficiency $\eta$, defined as the ratio of the total electron
%energy to the pulsar spin-down luminosity, is reasonable. 

We find that the fast-diffusion scenario can give an acceptable fit to the
$\gamma$-ray profile of the Geminga halo, which is in agreement with the result
by \citet{Recchia:2021kty}. In relativistic diffusion, the propagation of
newly injected electrons is ballistic, leading to a steep $\gamma$-ray profile
near the source similar to the measurement. However, the required $\eta$ is
150\%, which is too large. For LHAASO J0621$+$3755, the
fast diffusion scenario cannot give reasonable fit to the profile.
%The $\gamma$-ray profile predicted by the fast diffusion scenario is unable to
%fit the measurement after the convolution with the PSF. 
Therefore, the fast-diffusion scenario is strongly disfavored even with a relativistic correction.

In comparison, the slow-diffusion scenario can well fit the profiles with
reasonable $\eta$ for both the two halos. Furthermore, the slow-diffusion assumption also matches
the symmetry of the Geminga halo. If the diffusion coefficient is as large as the
Galactic average, the mean free path of electrons at 100 TeV reaches tens of parsecs,
which very likely corresponds to strong asymmetries of the halo
\citep{Lopez-Coto:2017pbk}. All these indicate that slow diffusion is still
necessary to interpret pulsar halos.

%slow diffusion is necessary

\appendix

\section{Discussion on the extreme cases of generalized J\"uttner propagator}
\label{sec:discuss}

In the case of $\lambda \gg ct$, we have $\kappa \ll 1$ and
\begin{equation}
 \frac{\kappa}{K_1(\kappa)}\ \approx \kappa^2\ .
\label{eq:ball}
\end{equation}
Substituting Eq.(\ref{eq:ball}) into Eq.(\ref{eq:P_rela}), we have
\begin{equation}
P_{\rm rela}(E,r,t) \approx \frac{1}{4\uppi(ct)^3}\ \frac{H(1-\xi)}{(1-\xi^2)^2}\ \kappa^2\ {\rm exp}\left(-\frac{\kappa}{\sqrt{1-\xi^2}}\right)\ = \frac{H(1-\xi)}{4\uppi(ct)^3}\ \frac{y{\rm d}y}{\xi{\rm d}\xi}\ {\rm exp}(-y)\ ,
\label{eq:appball}
\end{equation}
where $y = \kappa/\sqrt{1-\xi^2}$.

The ballistic propagator is written as
\begin{equation}
P_{\rm ball}(E,r,t) = \frac{1}{4\uppi c^3t^2}\ \delta(ct-r)\ .
\label{eq:P_ball}
\end{equation}
By evaluating the integral of Eq.(\ref{eq:P_ball}) multiplied by any function $q$ of $t$, we have
\begin{equation}
\int_{-\infty}^{+\infty} G_{\rm ball}(E,r,t)\ q(t)\ {\rm d}t = \frac{1}{4\uppi cr^2}\ q\left(\frac{r}{c}\right)\ .
\label{eq:intP_ball}
\end{equation}
Applying the same operation to Eq.(\ref{eq:appball}), we obtain
\begin{equation}
\begin{aligned}
\int_{-\infty}^{+\infty} P_{\rm rela}(E,r,t)\ q(t)\ {\rm d}t & \approx \int_{\frac{r}{c}}^{+\infty} \frac{1}{4\uppi(ct)^3}\ \frac{y{\rm d}y}{\xi{\rm d}\xi}\ {\rm exp}(-y)\ q(t)\ {\rm d}t\\ & = \int_{0}^{+\infty} \frac{1}{4\uppi(ct)^3}\ \frac{r}{c\xi^3}\ y\ {\rm exp}(-y)\ q[t(y)]\ {\rm d}y\\ & = \frac{1}{4\uppi cr^2}\ \int_{0}^{+\infty} y\ {\rm exp}(-y)\ q[t(y)]\ {\rm d}y\ ,
\end{aligned}
\end{equation}
which coincides with Eq.(\ref{eq:intP_ball}) because $\int_{0}^{+\infty} y\ {\rm exp}(-y)\ {\rm d}y = 1$ and the approximation
\begin{equation}
 t(y) = \frac{x}{c\sqrt{1-(\frac{\kappa}{y})^2}} \approx \frac{r}{c}\
\end{equation}
holds on the main part of integral interval.

In the case of $\lambda \ll ct$, we have $\kappa \gg 1$ and
\begin{equation}
 \frac{\kappa}{K_1(\kappa)} \approx \sqrt{\frac{2}{\uppi}}\ \kappa^{3/2}\ {\rm exp}(\kappa)\ .
\label{eq:diff}
\end{equation}
At a distance $r \lesssim \lambda$ where the electron number density is
non-ignorable, we obtain
\begin{equation}
 \xi = \frac{r}{ct} \lesssim \frac{\lambda}{ct} \ll 1.
\end{equation}
Substituting Eq.(\ref{eq:diff}) and the first-order approximations
$1/(1-\xi^2)^2 \approx 1$ and ${\rm exp}(-\kappa/\sqrt{1-\xi^2}) \approx {\rm exp}[-\kappa(1+\xi^2/2)]$ into Eq.(\ref{eq:P_rela}), we have
\begin{equation}
P_{\rm rela}(E,r,t) \approx \frac{1}{(ct)^3}\ \frac{1}{(1-\xi^2)^2}\ \left(\frac{\kappa}{2\uppi}\right)^{3/2}\ {\rm exp}\left(-\frac{\kappa\xi^2}{2}\right)\ = \frac{1}{(\uppi\lambda^2)^{3/2}}\ {\rm exp} \left(-\frac{r^2}{\lambda^2}\right)\ ,
\end{equation}
which coincides with Eq.(\ref{eq:P_diff}), indicating that the propagator
transits to the normal one.

\section{Discussion on the small-angle diffusion approximation}
\label{sec:M_alpha}

\begin{figure}
	\centering
	\includegraphics[width=5cm]{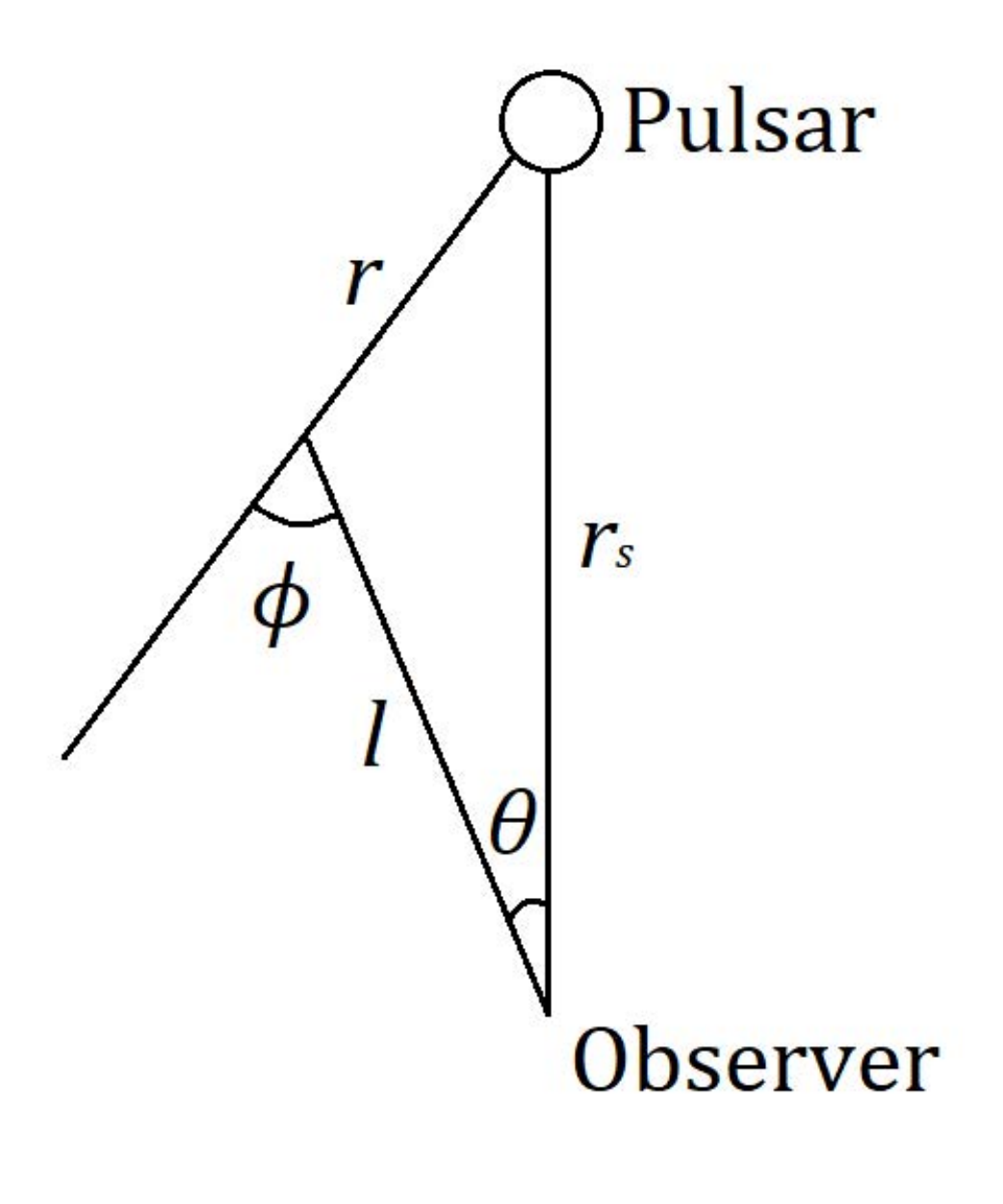}
    \includegraphics[width=8cm]{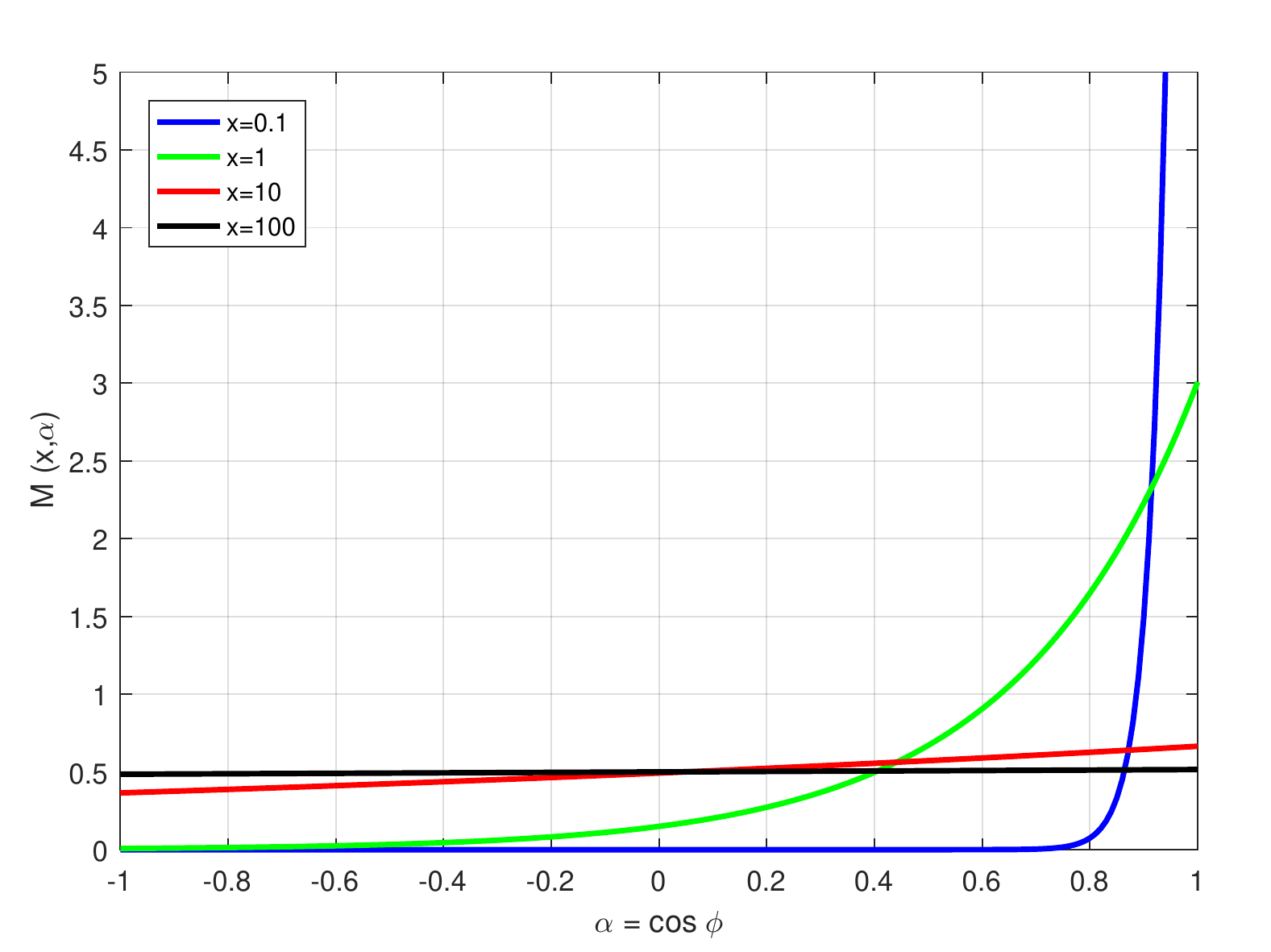}
	\caption{Left: A sketch map for the geometric parameters. Right: The small-angle diffusion approximation $M$ as a function of $\alpha$ for different $x$.}
	\label{fig:M_alpha}
\end{figure}

For the pure diffusion model, we directly integrate the electron number density $N$
over the line of sight to get the electron surface density
\begin{equation}
F_{\mathrm{diff}}(E,\theta) = \int^{\infty}_{0} N(E,l,\theta)\ \mathrm{d}l\ ,
\end{equation}
and then apply the standard ICS calculation, because the angular distribution of
$\gamma$-ray is isotropic. But for the corrective model we use in this work, the angular
distribution of $\gamma$-ray emitted by the quasi-ballistic electrons within the vicinity
of source is anisotropic. Therefore, a small-angle correction is needed.

As a result of relativistic beaming effect, the angular distribution of $\gamma$-ray can be
regarded as the velocity angular distribution of electrons.
For the electrons with energy $E$ at the distance of $r$ from the source, let
$\mathcal{M}(E,r,\Omega)\ \mathrm{d}\Omega$ represent the proportion that moves towards
$\Omega(\phi,\varphi)$ within the solid angle $\mathrm{d}\Omega$. Here $\phi$ is the angle
between the radial direction and the line of sight, as shown in the left panel of
Fig.~\ref{fig:M_alpha} together with other geometric parameters. From geometry, we derive
$r(l,\theta) = \sqrt{r^{2}_{s}+l^2-2r_{s}l\mathrm{cos}\theta}$ and
$\alpha(l,\theta) = \mathrm{cos}(\phi) = (r_s\mathrm{cos}\theta-l)/r$.

We introduce a dimensionless parameter $x(E,r) = rc/D(E)$. Because of the symmetry around
the radial direction, the angle $\varphi$ is arbitrary and can be removed by integral:
\begin{equation}
 \mathcal{M}(E,r,\Omega) = \mathcal{M}(x,\phi,\varphi) = \frac{1}{2\uppi}\ M(x,\phi).
\end{equation}
Taking the limit as $x \rightarrow \infty$, we get $\mathcal{M}_{\rm{diff}} = \frac{1}{4\uppi}$
and $M(\infty,\phi) = \frac{1}{2}$, because the case transits to pure diffusion scenario,
as shown in the right panel of Fig.~\ref{fig:M_alpha}. Generally, $\mathcal{M}(E,r,\Omega)$
should be normalized by the solid angle $\Omega$.
By $\mathrm{d}\Omega = \rm{sin}\phi\ \mathrm{d}\phi\mathrm{d}\varphi$, we derive
\begin{equation}
\begin{aligned}
 1 & = \oiint \mathcal{M}(E,r,\Omega)\ \mathrm{d}\Omega = \oiint \mathcal{M}(x,\phi,\varphi)\ \rm{sin}\phi\ \mathrm{d}\phi\mathrm{d}\varphi\\ & =\int^{2\uppi}_{0}\ \mathrm{d}\varphi\ \int^{\uppi}_{0}\frac{1}{2\uppi}\ M(x,\phi)\ \mathrm{sin}\phi\ \rm{d}\phi\\ & = \int^{-1}_{1} M(x,\phi)\ \mathrm{d}(-\mathrm{cos}\phi) = \int^{1}_{-1} M(x,\alpha)\ \mathrm{d}\alpha\ ,
\end{aligned}
\end{equation}
where $\alpha = \mathrm{\phi}$. \citet{Prosekin:2015} gives the form
$M \sim \mathrm{exp}[-\frac{3(1-\alpha)}{x}]$, and the normalization function is
\begin{equation}
 Z(x) = \int^{1}_{-1} \mathrm{exp}[-\frac{3(1-\alpha)}{x}]\ \mathrm{d}\alpha\ = \frac{x}{3}[1-\mathrm{exp}(-\frac{6}{x})]\ .
\end{equation}
This is exactly Eq.~(\ref{eq:M_alpha}) we apply to our calculation. Then we get the apparent
electron surface density:
\begin{equation}
 F(E,\theta) = \int^{\infty}_{0} N(E,l,\theta)\ \frac{\mathcal{M}(E,r,\Omega)}{\mathcal{M}_{\rm{diff}}}\ \mathrm{d}l = \int^{\infty}_{0} 2N(E,l,\theta)M(x,\alpha)\ \mathrm{d}l\ ,
\end{equation}
which is equivalent to the pure diffusion model and can be directly used in the standard ICS calculation.

\section{Discussion on the cutoff energy and the power-law index of the injection spectrum}
\label{sec:p_eta}

 \begin{figure}
 	\centering
 	\includegraphics[width=10cm]{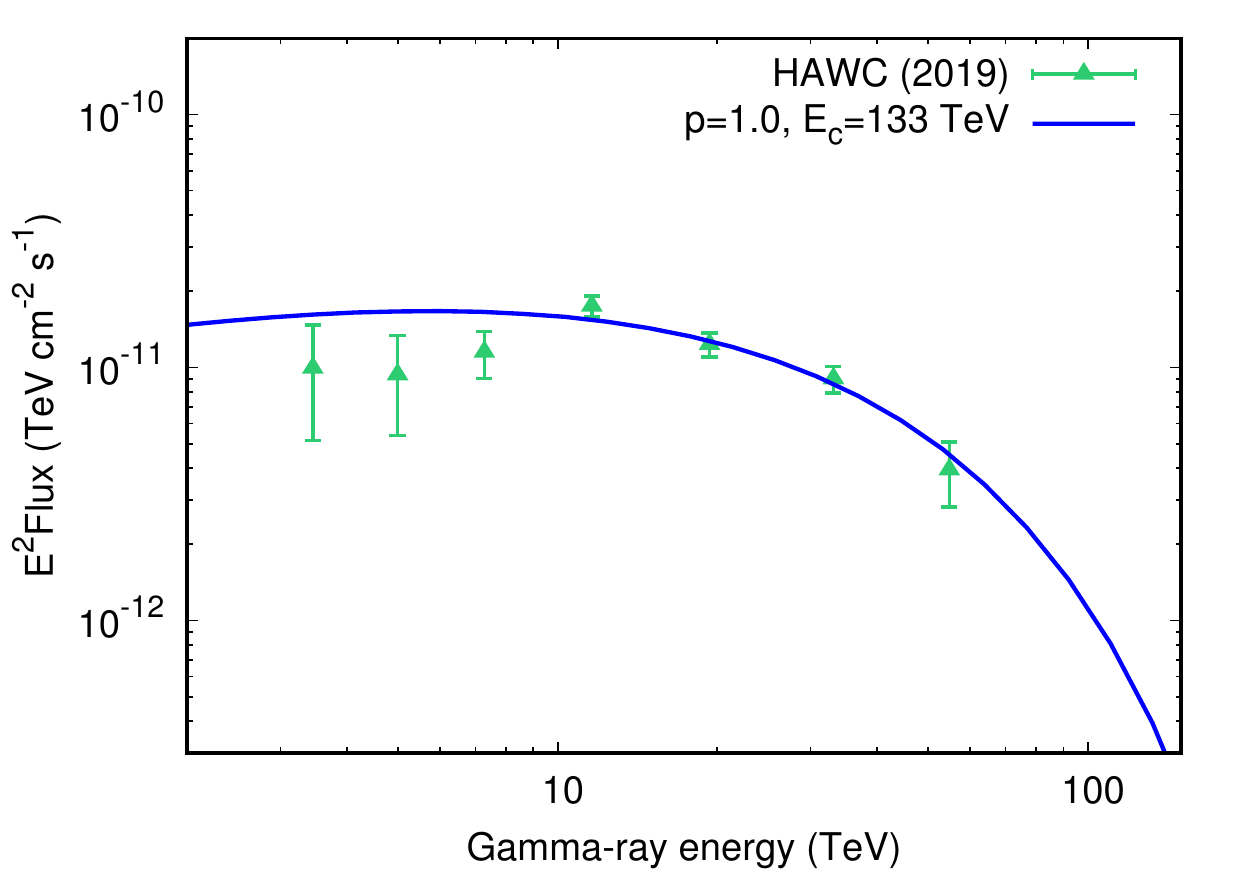}
 	\caption{ The best-fit $\gamma$-ray spectrum to the HAWC data. }
 	\label{fig:spec}
\end{figure}

\begin{figure}
	\centering
	\includegraphics[width=10cm]{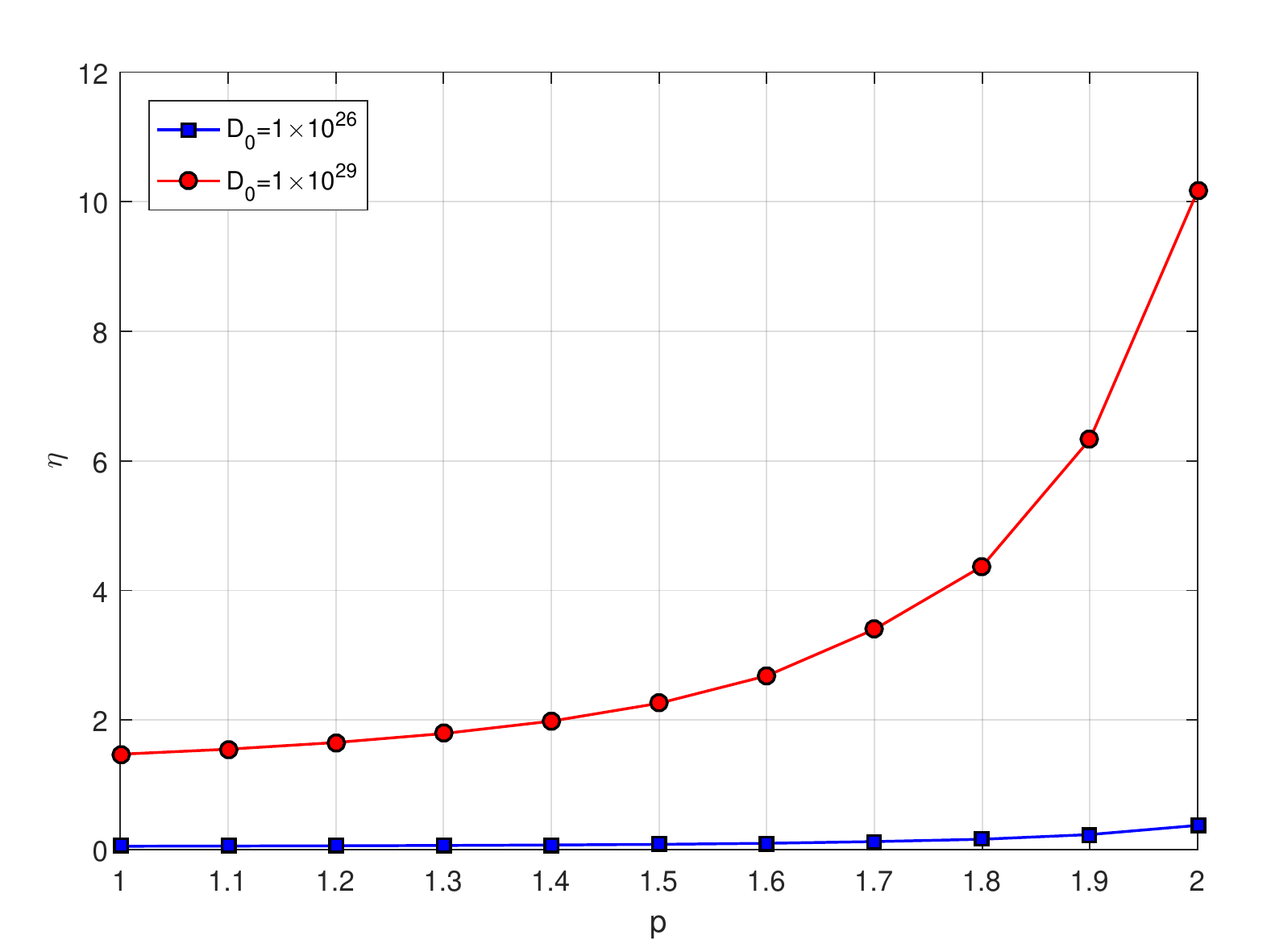}
	\caption{ The relation between the power-law index and the required conversion 
efficiency to fit the Geminga profile. }
	\label{fig:p_eta}
\end{figure}

The preliminary $\gamma$-ray spectrum of HAWC indicates that the electron
spectrum cannot be a simple power law, but more likely a power law with a
high-energy cutoff \citep{2019ICRC...36..832Z}. This is also supported by
the GeV observation of the Geminga halo \citep{xi2019gev}. 
%The cutoff energy $E_c$ used in the main text is determined by
%a fit to the preliminary spectrum of HAWC.
However, the HAWC $\gamma$-ray spectrum is not broad enough to
determine the power-law index $p$. The value of $p=1.0$ is taken from the
observation of the X-ray PWN of Geminga \citep{pavlov2006pulsar}.
As PWNe can be seen as the sources of the parent electrons of pulsar halos, the
X-ray spectrum of PWNe could be used to estimate the low-energy part of the
electron injection spectrum. The X-ray PWN of Geminga has a hard photon spectrum
with an index of 1.0, corresponding to $p=1.0$ for the electron spectrum. 

We give a careful fit to the preliminary $\gamma$-ray spectrum of HAWC
\citep{2019ICRC...36..832Z} to determine the injection spectrum. The best-fit cutoff
energy $E_c$ is 133~TeV. The spectral index $p$ is fixed to be 1.0 as suggested
by the observations of Geminga's X-ray PWN. The best-fit $\gamma$-ray spectrum to the
HAWC data is shown in Fig.~\ref{fig:spec}.

Furthermore, the power-law indices of PWNe are generally smaller than 2.0 (e.g,,
$p=1.8$ for Vela X). The GeV observation of the Geminga halo also indicates
$p<2.0$ \citep{xi2019gev}. For $p<2.0$, the integral energy of the
electron spectrum is concentrated at the high-energy end of the spectrum, which
is not seriously affected by the choice of $p$.

In Fig.~\ref{fig:p_eta}, we show the relation between the power-law index $p$ and the
required conversion efficiency $\eta$ to best fit the $\gamma$-ray profile for
both the slow-diffusion and fast-diffusion cases.
It can be seen that for $p<2.0$, $\eta$ never exceed 100\% for the slow-diffusion case,
while $\eta$ is always larger than 100\% for the fast-diffusion case. Thus, our
conclusion is not affected by the assumption of the injection spectrum.

\begin{acknowledgements}
This work is supported by the National Natural Science Foundation of China under 
grant Nos. 12175248 and 12105292.
\end{acknowledgements}

\bibliography{refs}{}

\begin{thebibliography}{}
\expandafter\ifx\csname natexlab\endcsname\relax\def\natexlab#1{#1}\fi
\providecommand{\url}[1]{\href{#1}{#1}}
\providecommand{\dodoi}[1]{doi:~\href{http://doi.org/#1}{\nolinkurl{#1}}}
\providecommand{\doeprint}[1]{\href{http://ascl.net/#1}{\nolinkurl{http://ascl.net/#1}}}
\providecommand{\doarXiv}[1]{\href{https://arxiv.org/abs/#1}{\nolinkurl{https://arxiv.org/abs/#1}}}

\bibitem[{Abeysekara {et~al.}(2017)Abeysekara, Albert, Alfaro, Alvarez,
  Álvarez, Arceo, Arteaga-Velázquez, Rojas, Solares, Barber, Bautista-Elivar,
  Becerril, Belmont-Moreno, BenZvi, Berley, Bernal, Braun, Brisbois,
  Caballero-Mora, Capistrán, Carramiñana, Casanova, Castillo, Cotti, Cotzomi,
  de~León, León, la~Fuente, Dingus, DuVernois, Díaz-Vélez, Ellsworth,
  Engel, Enríquez-Rivera, Fiorino, Fraija, García-González, Garfias,
  Gerhardt, Muñoz, González, Goodman, Hampel-Arias, Harding, Hernández,
  Hernández-Almada, Hinton, Hona, Hui, Hüntemeyer, Iriarte, Jardin-Blicq,
  Joshi, Kaufmann, Kieda, Lara, Lauer, Lee, Lennarz, Vargas, Linnemann,
  Longinotti, Raya, Luna-García, López-Coto, Malone, Marinelli, Martinez,
  Martinez-Castellanos, Martínez-Castro, Martínez-Huerta, Matthews,
  Miranda-Romagnoli, Moreno, Mostafá, Nellen, Newbold, Nisa, Noriega-Papaqui,
  Pelayo, Pretz, Pérez-Pérez, Ren, Rho, Rivière, Rosa-González, Rosenberg,
  Ruiz-Velasco, Salazar, Greus, Sandoval, Schneider, Schoorlemmer, Sinnis,
  Smith, Springer, Surajbali, Taboada, Tibolla, Tollefson, Torres, Ukwatta,
  Vianello, Weisgarber, Westerhoff, Wisher, Wood, Yapici, Yodh, Younk, Zepeda,
  Zhou, Guo, Hahn, Li, \& Zhang}]{Abeysekara:2017old}
Abeysekara, A.~U., Albert, A., Alfaro, R., {et~al.} 2017, Science, 358, 911,
  \dodoi{10.1126/science.aan4880}

\bibitem[{Aharonian {et~al.}(2021)Aharonian, An, Axikegu, Bai, Bai, Bao,
  Bastieri, Bi, Bi, Cai, Cai, Cao, Cao, Chang, Chang, Chang, Chen, Chen, Chen,
  Chen, Chen, Chen, Chen, Chen, Chen, Chen, Chen, Chen, Chen, Cheng, Cheng,
  Cui, Cui, Cui, Dai, Dai, Dai, Danzengluobu, della Volpe, D'Ettorre~Piazzoli,
  Dong, Fan, Fan, Fan, Fang, Fang, Feng, Feng, Feng, Feng, Gao, Gao, Gao, Gao,
  Ge, Geng, Gong, Gou, Gu, Guo, Guo, Guo, Guo, Han, He, He, He, He, He, He,
  Heller, Hor, Hou, Hou, Hu, Hu, Hu, Hu, Huang, Huang, Huang, Huang, Huang, Ji,
  Ji, Jia, Jiang, Jiang, Jin, Kuleshov, Levochkin, Li, Li, Li, Li, Li, Li, Li,
  Li, Li, Li, Li, Li, Li, Li, Li, Li, Li, Liang, Liang, Lin, Liu, Liu, Liu,
  Liu, Liu, Liu, Liu, Liu, Liu, Liu, Liu, Liu, Liu, Liu, Liu, Long, Lu, Lv, Ma,
  Ma, Ma, Mao, Masood, Mitthumsiri, Montaruli, Nan, Pang, Pattarakijwanich,
  Pei, Qi, Ruffolo, Rulev, S\'aiz, Shao, Shchegolev, Sheng, Shi, Song, Stenkin,
  Stepanov, Sun, Sun, Sun, Tam, Tang, Tian, Wang, Wang, Wang, Wang, Wang, Wang,
  Wang, Wang, Wang, Wang, Wang, Wang, Wang, Wang, Wang, Wang, Wang, Wang, Wang,
  Wang, Wang, Wei, Wei, Wei, Wen, Wu, Wu, Wu, Wu, Wu, Xi, Xia, Xia, Xiang,
  Xiao, Xiao, Xin, Xin, Xing, Xu, Xu, Xue, Yan, Yang, Yang, Yang, Yang, Yang,
  Yang, Yang, Yao, Yao, Ye, Yin, Yin, You, You, Yu, Yuan, Zeng, Zeng, Zeng,
  Zeng, Zha, Zhai, Zhang, Zhang, Zhang, Zhang, Zhang, Zhang, Zhang, Zhang,
  Zhang, Zhang, Zhang, Zhang, Zhang, Zhang, Zhang, Zhang, Zhang, Zhang, Zhang,
  Zhao, Zhao, Zhao, Zhao, Zhao, Zheng, Zheng, Zhou, Zhou, Zhou, Zhou, Zhou,
  Zhou, Zhu, Zhu, Zhu, Zhu, Zuo, \& Huang}]{Aharonian2021extended}
Aharonian, F., An, Q., Axikegu, {et~al.} 2021, Phys. Rev. Lett., 126, 241103,
  \dodoi{10.1103/PhysRevLett.126.241103}

\bibitem[{Aloisio {et~al.}(2009)Aloisio, Berezinsky, \&
  Gazizov}]{Aloisio:2008tx}
Aloisio, R., Berezinsky, V., \& Gazizov, A. 2009, Astrophys. J., 693, 1275,
  \dodoi{10.1088/0004-637x/693/2/1275}

\bibitem[{Blumenthal \& Gould(1970)}]{Blumenthal:1970gc}
Blumenthal, G.~R., \& Gould, R.~J. 1970, Rev. Mod. Phys., 42, 237,
  \dodoi{10.1103/RevModPhys.42.237}

\bibitem[{Dunkel {et~al.}(2007{\natexlab{a}})Dunkel, Talkner, \&
  H\"anggi}]{Dunkel:2007relativistic}
Dunkel, J., Talkner, P., \& H\"anggi, P. 2007{\natexlab{a}}, Phys. Rev. D, 75,
  043001, \dodoi{10.1103/PhysRevD.75.043001}

\bibitem[{Dunkel {et~al.}(2007{\natexlab{b}})Dunkel, Talkner, \&
  Hänggi}]{dunkel2007relative}
Dunkel, J., Talkner, P., \& Hänggi, P. 2007{\natexlab{b}}, New J. Phys., 9,
  144, \dodoi{10.1088/1367-2630/9/5/144}

\bibitem[{Fang {et~al.}(2021{\natexlab{a}})Fang, Bi, Lin, \&
  Yuan}]{Fang:2020dmi}
Fang, K., Bi, X.-J., Lin, S.-J., \& Yuan, Q. 2021{\natexlab{a}}, Chin. Phys.
  Lett., 38, 039801, \dodoi{10.1088/0256-307x/38/3/039801}

\bibitem[{Fang {et~al.}(2021{\natexlab{b}})Fang, Xi, \& Bi}]{Fang:2021qon}
Fang, K., Xi, S.-Q., \& Bi, X.-J. 2021{\natexlab{b}}, Phys. Rev. D, 104,
  103024, \dodoi{10.1103/PhysRevD.104.103024}

\bibitem[{J{\"u}ttner(1911)}]{Juttner1911maxwellsche}
J{\"u}ttner, F. 1911, Annalen der Physik, 339, 856,
  \dodoi{https://doi.org/10.1002/andp.19113390503}

\bibitem[{Linden {et~al.}(2017)Linden, Auchettl, Bramante, Cholis, Fang,
  Hooper, Karwal, \& Li}]{Linden:2017vvb}
Linden, T., Auchettl, K., Bramante, J., {et~al.} 2017, Phys. Rev. D, 96,
  103016, \dodoi{10.1103/PhysRevD.96.103016}

\bibitem[{López-Coto \& Giacinti(2018)}]{Lopez-Coto:2017pbk}
López-Coto, R., \& Giacinti, G. 2018, Mon. Not. Roy. Astron. Soc., 479, 4526,
  \dodoi{10.1093/mnras/sty1821}

\bibitem[{Pavlov {et~al.}(2006)Pavlov, Sanwal, \& Zavlin}]{pavlov2006pulsar}
Pavlov, G.~G., Sanwal, D., \& Zavlin, V.~E. 2006, Astrophys. J., 643, 1146,
  \dodoi{10.1086/503250}

\bibitem[{Prosekin {et~al.}(2015)Prosekin, Kelner, \&
  Aharonian}]{Prosekin:2015}
Prosekin, A.~Y., Kelner, S.~R., \& Aharonian, F.~A. 2015, Phys. Rev. D, 92,
  083003, \dodoi{10.1103/PhysRevD.92.083003}

\bibitem[{Recchia {et~al.}(2021)Recchia, Di~Mauro, Aharonian, Orusa, Donato,
  Gabici, \& Manconi}]{Recchia:2021kty}
Recchia, S., Di~Mauro, M., Aharonian, F.~A., {et~al.} 2021, Phys. Rev. D, 104,
  123017, \dodoi{10.1103/PhysRevD.104.123017}

\bibitem[{Xi {et~al.}(2019)Xi, Liu, Huang, Fang, \& Wang}]{xi2019gev}
Xi, S.-Q., Liu, R.-Y., Huang, Z.-Q., Fang, K., \& Wang, X.-Y. 2019, Astrophys.
  J., 878, 104, \dodoi{10.3847/1538-4357/ab20c9}

\bibitem[{Yuan {et~al.}(2017)Yuan, Lin, Fang, \& Bi}]{Yuan:2017ozr}
Yuan, Q., Lin, S.-J., Fang, K., \& Bi, X.-J. 2017, Phys. Rev. D, 95, 083007,
  \dodoi{10.1103/PhysRevD.95.083007}

\bibitem[{{Zhou}(2019)}]{2019ICRC...36..832Z}
{Zhou}, H. 2019, in International Cosmic Ray Conference, Vol.~36, 36th
  International Cosmic Ray Conference (ICRC2019), 832

\end{thebibliography}
\bibliographystyle{aasjournal}

\end{document}